\documentstyle[editedvolume]{jd}
\input psfig

\begin{opening}
\title{Disk Mass from Large-scale Dynamics}

\author{J.\ A.\ Sellwood}
\institute{Rutgers University\\
         Dept.\ of Physics \& Astronomy, PO Box 849, Piscataway, NJ 08855, USA}

\end{opening}

\runningtitle{Disk Mass}

\begin{document}
\noindent Rutgers Astrophysics preprint \# 225.

\noindent To appear in {\it Highlights of Astronomy\/} {\bf 11}, ed.\ J.\ Andersen (Dordrecht: Kluwer Academic Publishers)

\section{Introduction}
The radial distribution of mass in a disk galaxy is strongly constrained by its 
rotation curve.  The separate contributions from the individual stellar 
populations and dark matter (DM) are not easily disentangled, however, 
especially since there is generally no feature to indicate where the component 
dominating the central attraction switches from luminous to dark matter.  Here I 
summarize three recent thesis projects at Rutgers University which all suggest 
that DM has a low density in the inner parts of bright galaxies, and that most 
of the mass therefore resides in the disk.  In addition, I present some 
preliminary work on the Milky Way.  If we are able to determine the M/L of a 
typical disk stellar population, it should provide a useful constraint on the 
numbers of low mass stars.

\section{Masses of galactic disks}
Palunas \& Williams (1997) have I-band surface photometry and 2-D kinematic maps 
of 76 southern disk galaxies.  They calculate the gravitational field in each 
case, assuming a constant M/L for the disk and bulge, and find that the shape of 
the inner rotation curve is very well reproduced by the light distribution for 
M/L$_{\rm I} \sim 2-3$ in most cases (for $H_0=75$ km s$^{-1}$ Mpc$^{-1}$).  
Mass discrepancies indicative of the dark halo contribution become noticeable 
only in the outer parts of some galaxies.

Weiner (1997) models the 2-D flow pattern of gas in the barred galaxy NGC 4123 
by calculating gas flows in a model potential derived from the I-band 
photometry.  He finds that a M/L$_{\rm I} \sim 2.5$ is again required for the 
predicted flow pattern to match that observed.  As a result, almost all the mass 
in the inner galaxy must be distributed as the luminous component, else the 
rotation curve in the outer disk would be too high.

Debattista \& Sellwood (1997) use fully self-consistent simulations of barred 
disks embedded in halos to verify Weinberg's (1985) prediction that the rotation 
rate of the bar should quickly decrease through dynamical friction.  In even a 
moderate density halo, the bar slows dramatically as it loses angular momentum 
and the ratio of corotation radius to bar semi-major axis quickly becomes 
inconsistent with observed properties of barred galaxies.  Only if the central 
halo density is low, and the disk is massive, can rapid braking be avoided.

\section{The Milky Way}
Both Spergel et al.\ (1996) and Freudenreich (1997) have proposed models for the 
3-D luminosity density in the Milky Way that are based on the all-sky 
multi-color photometry obtained by the DIRBE instrument on the COBE satellite.  
Both groups take extinction by dust into account, but in different ways, and 
then fit parameters of an idealised model to the projected sky brightness.  
While the volume luminosity density distributions are different in the two 
models, the integrated luminosities for the Milky Way in the different IR color 
bands (see Mulhotra et al.\ 1996) are very similar.  Freudenreich gives a useful 
conversion to bolometric luminosity for the whole galaxy: L$_{\rm bol} = 2.3 
\times 10^{10}$ L$_\odot$~($\pm 20$\%).

We can use these models to derive a mass density for the light distributions, 
provided a constant M/L assumption is valid.  I make this assumption in the 
present work, but it should be borne in mind that localized excesses of 
supergiant stars may cause features in the IR light distribution that have no 
counterparts in the mass; one such feature may be the IR bright ring at 3 kpc 
where the molecular gas density is high, which could be responsible for the 
apparently short scale length for the disk.  It is also questionable whether the 
disk and bulge/bar should be assigned the same M/L.

I determine my single free parameter, the M/L for the Milky Way, by requiring 
that the surface density of luminous mass at the Sun be approximately 50 
M$_\odot$ pc$^{-2}$, in accordance with recent determinations (Kuijken \& 
Gilmore 1991).  A safer normalization can be made by modelling the gas flow in 
the barred region (Englmaier \& Gerhard, this meeting) and it is encouraging 
that they also find that the mass of the inner galaxy is dominated by luminous 
material.

\begin{figure}[t]
\psfig{figure=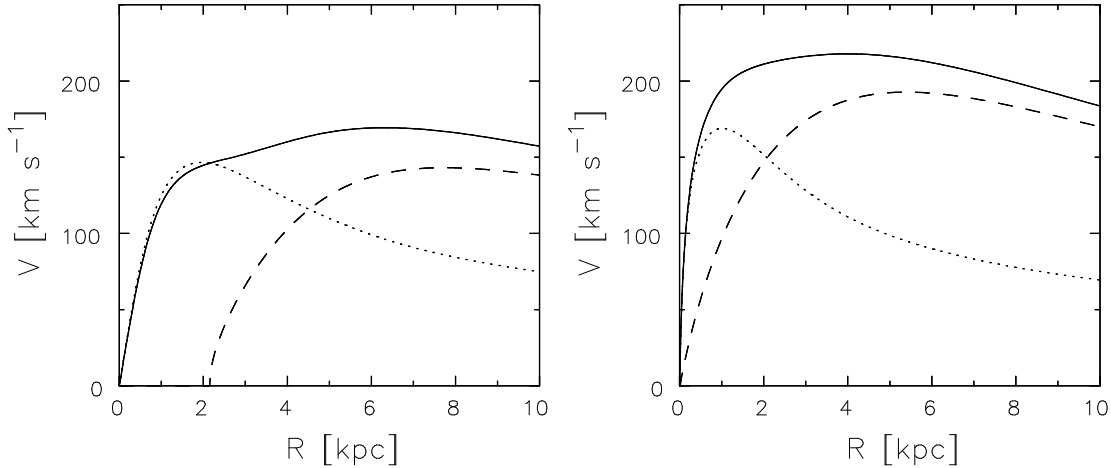,width=0.9\hsize,angle=0}
\caption{Axially symmetrised ``rotation curves'' derived from the photometric 
models of the Milky Way proposed by Freudenreich (left) with M/L$_{\rm bol}=2$ 
and Spergel et al.\ (right) M/L$_{\rm bol}=3$.  The dotted (dashed) curve shows 
the bar/bulge (disk) contributions, the full curve is their sum.}
\end{figure}

Both models have two stellar components, the bulge/bar and a disk.  The rotation 
curve obtained by solving for the gravitational field is shown in Figure 1; the 
inner ``rotation curve'' of these barred galaxy models is simply derived from 
the azimuthal average of the central attraction.  Freudenreich's model S has a 
hole in the disk; setting M/L$_{\rm bol}=2$ gives a surface density of 46 
M$_\odot$ pc$^{-2}$ at $R=8$ kpc.  The model preferred by Spergel et al.\ has 
not yet been published, but Binney, Gerhard \& Spergel (1997) give one fit which 
has almost the same total light as Freudenreich's, even though the disk in this 
model does not have a hole.  The circular velocity curve shown in Figure 1(b) 
results from setting M/L$_{\rm bol}=3$ to give a local surface density of 49 
M$_\odot$ pc$^{-2}$.  The central brightness, extrapolated inwards from 3 kpc, 
is very high in this model and clearly overstates the disk light in this region; 
a greater fraction of the light probably should be attributed to the bar.

\section{Conclusions}
Both studies of external galaxies, as well as of the Milk Way, suggest that the 
disk component contains most of the mass in the bright inner parts of high 
luminosity galaxies.  The data seem to suggest a M/L$_{\rm I}\sim 2-3$ for the 
stellar population in typical galaxy disk.  Very preliminary fits of constant 
M/L models for the Milky Way suggest M/L$_{\rm bol}\sim 2-3$ (a value that also 
indicates little DM in the inner Galaxy).

\smallskip
\noindent This work was supported by NSF grant AST 93/18617 and NASA Theory 
grant NAG 5-2803.

%\section{References}\parindent=0pt

\end{document}